\documentclass[a4paper, aps, prl, reprint, superscriptaddress]{revtex4-1}

\usepackage{graphicx}
\usepackage{amsmath}
\usepackage{amssymb}
\usepackage{color}
\newcommand{\sout}[1]{}
\newcommand{\unit}[1]{~\mathrm{#1}}
\newcommand{\unita}[1]{\mathrm{#1}}

\newcommand{\Rsq}{R_\Box}
\newcommand{\Tc}{T_\mathrm{c}}
\newcommand{\kB}{k_\mathrm{B}}
\newcommand{\Tco}{T_\mathrm{c0}}
\newcommand{\red}[1]{#1}
\usepackage{epstopdf}
\begin{document}

\title{Strongly disordered TiN and NbTiN $s$-wave superconductors probed by microwave electrodynamics}

\author{E. F. C. Driessen}
\email{e.f.c.driessen@tudelft.nl}
\affiliation{Kavli Institute of Nanoscience, Delft University of Technology, Lorentzweg 1, 2628 CJ Delft, The Netherlands}

\author{P. C. J. J. Coumou}
\affiliation{Kavli Institute of Nanoscience, Delft University of Technology, Lorentzweg 1, 2628 CJ Delft, The Netherlands}

\author{R. R. Tromp}
\affiliation{Kavli Institute of Nanoscience, Delft University of Technology, Lorentzweg 1, 2628 CJ Delft, The Netherlands}

\author{P. J. de Visser}
\affiliation{Kavli Institute of Nanoscience, Delft University of Technology, Lorentzweg 1, 2628 CJ Delft, The Netherlands}
\affiliation{SRON National Institute for Space Research, Sorbonnelaan 2, 3584 CA Utrecht, The Netherlands}

\author{T. M. Klapwijk}
\affiliation{Kavli Institute of Nanoscience, Delft University of Technology, Lorentzweg 1, 2628 CJ Delft, The Netherlands}

\begin{abstract}
We probe the effects of strong disorder ($2.4<k_\mathrm{F}l<8.6$) on superconductivity in thin films of niobium titanium nitride and titanium nitride by measuring the microwave electrodynamics in coplanar waveguide resonators. We find a gradual evolution of the electromagnetic response with disorder, deviating from BCS theory, for both materials. This result can be understood as due to changes in the quasiparticle density of states, induced by the short elastic scattering length. The observations are consistent with a model using an effective pair breaker, dependent on the level of disorder.
\end{abstract}

\maketitle

It has been a longstanding paradigm in superconductivity, that the properties of the superconducting state are not affected by disorder~\cite{Anderson:1959ub}. Consequently, it has become a justified practice to expect a fixed critical temperature $\Tc$ for superconducting films, directly related to a standard BCS gap and quasiparticle density of states, irrespective of  their large differences in normal state resistivity.
On the other hand, theoretical evidence has been developed, that reveals severe deviations from BCS theory for materials with a resistivity in the range $100\unit{\mu\Omega cm}$ and higher. In materials with such a large resistivity, the elastic scattering length $l$ is of the order of the interatomic distance. Therefore, it can be expected that localization effects become important. From numerical simulations it has become clear that - even for homogeneous disorder - eventually an inhomogeneous superconducting state will arise, when the disorder is increased~\cite{Ghosal:1998up}. A short elastic scattering length enhances both the Coulomb interaction between electrons~\cite{Finkelshtein:1987tl}, and the  interference of electrons scattering from impurities. These mesoscopic fluctuations were shown to grow when approaching the superconductor-to-insulator transition~\cite{Skvortsov:2005eb}, and unavoidably affect the properties of the superconducting state~\cite{Feigelman:2011vj}.

In early tunneling experiments, it was shown that in films of granular aluminum the coherence peak in the quasiparticle density of states  gets smeared out significantly when the amount of disorder is increased~\cite{DYNES:1984wu}. More recently, it was found that increasing the level of disorder leads to a direct superconductor-to-insulator transition for various materials~\cite{Gantmakher:2010il}. It was found in thin TiN films that, close to this transition, the quasiparticle gap shows large spatial fluctuations~\cite{Sacepe:2008jx}, and by similar tunneling experiments in InO$_x$, evidence was obtained that on the insulating side of the superconductor-to-insulator transition, there is a peculiar gapped state, presumably containing localized Cooper pairs~\cite{Sacepe:2011jm}.

From both experimental and theoretical points of view it has thus become clear that superconducting films with a large resistivity, are unavoidably prone to deviations from BCS theory, even when \red{\sout{they are perfectly homogeneous} the disorder leading to the large resistivity is homogeneously distributed}.
Proper understanding of these intrinsic deviations is interesting from a fundamental point of view, as well as crucial for a variety of experiments:  the interest in highly resistive superconducting films has taken a big leap over the last years in the development of sensitive detectors~\cite{LeDuc:2010do}, and in experiments involving quantum phase slips~\cite{Astafiev:2012bi}, or where a superconductor is used in combination with a large magnetic field~\cite{Popinciuc:2012tb,Mourik:2012je,Note1}
. In all these applications, a proper understanding of the superconducting state is necessary to interpret performance of the devices and the results of experiments, especially when the nature of the electronic states is an important experimental ingredient~\cite{Lutchyn:2011wk}.

Until now, most experimental effort in understanding the superconducting state at high levels of disorder has been focused on magnetoresistance measurements as a function of disorder, yielding information close to $\Tc$, or on position-dependent scanning tunneling spectroscopy, showing the local properties of the material. In this Letter, we use instead the response of a superconducting resonator at microwave frequencies as a probe for the global superconducting state 
at low temperatures $T\ll\Tc$, allowing the convenient study of films with a variety of resistivities. We study a series of titanium nitride and niobium titanium nitride films with varying normal-state resistivity, $120\unit{\mu\Omega cm}<\rho<550\unit{\mu\Omega cm}$, still away from the superconductor-to-insulator transition. For increasing disorder, we find increasing deviations from BCS theory.  We show that these deviations are closely linked to the short elastic scattering length in the films and compare them to the available theory for strongly disordered superconductors.


Table~\ref{tab:parameters} gives an overview of the eight films that are studied in this Letter. 
The series of TiN films (A-F) was grown using plasma-enhanced atomic layer deposition (ALD)~\cite{SUPINF}. This process is self-limiting, and therefore in principle allows for growing the TiN film one monolayer at a time. In practice, the average growth rate of the TiN film is 0.45~$\textrm{\AA}$ per cycle. The films are grown with a thickness ranging from 6~nm to 89~nm. The structure of the films is of a polycrystalline nature, as determined with atomic force microscopy. The grain size $b$ increases with increasing film thickness, and is significantly larger for film F, that was deposited on passivated silicon instead of on native silicon oxide. From the fact that the other parameters of film F do not differ from the trend observed in the other films, we infer that the grain boundaries are not limiting the electronic transport properties.
The NbTiN films G and H were deposited using DC magnetron sputtering from a Nb$_{0.7}$Ti$_{0.3}$ target. The resistivity was controlled by varying the pressure in the sputtering chamber during deposition~\cite{SUPINF}.

\begingroup
\begin{table}[t]
\caption{Parameters of the films studied in this experiment.}\label{tab:parameters}
\begin{tabular}{cccccccccc}
\hline\hline
Film	 &Sub-		& $d$	& $b$	& $\rho$			& $k_\mathrm{F}l$	& $l$ 	& $\tau$	& $\Tc$	& $\alpha$\\
ID	&strate		& [nm]	& [nm]	&[$\mu\Omega$cm]	& 	 			& [\AA]	& [fs]		& [K]		&[$\kB\Tc'$]\\
\hline
\multicolumn{10}{c}{\emph{TiN ALD-deposited films}}\\
A	&SiO$_2$		& 6	
						& 25		& 380			& 3.3		& 3.4		& 		& 1.5		& 0.22\\
B	&SiO$_2$		& 11		& 27		& 356 			& 3.5		& 3.5		& 1.2		& 2.2		& 0.17\\
C	&SiO$_2$		& 22		& 32		& 253			& 4.6		& 4.4		& 		& 2.7		& 0.13\\
D	&SiO$_2$		& 45		& 37		& 187			& 6.1		& 5.7		& 1.2		& 3.2		& 0.10\\
E	&SiO$_2$		& 89		& 42		& 120			& 8.6		& 7.3		& 1.7		& 3.6		& 0.01\\
F	&H-Si		& 55		& 44		& 212			& 6.0		& 6.4		& 1.4		& 3.3		& 0.08\\
\hline
\multicolumn{10}{c}{\emph{NbTiN sputter-deposited films}}\\
G	& Sapphire	& 300	& 85		& 150			& 8.2		& 6.3		& 		& 14.8	& 0.15\\
H	& H-Si		& 50		& 30		& 506			& 2.4		& 2.4		& 		& 11.9	& 0.34\\
\hline\hline
\end{tabular}
\end{table}
\endgroup

All free-electron parameters were determined at a temperature $T=10\unit{K}$ for the TiN films, and $T=40\unit{K}$ for the NbTiN films. Resistance, Hall effect, and upper critical field were determined on a Hall bar structure for each film.
From the measured carrier density and diffusion constant, values for the disorder parameter $k_\mathrm{F}l$, where $k_\mathrm{F}$ is the Fermi wave vector, and the elastic scattering time $\tau$ were determined using free-electron theory~\cite{SUPINF}. $k_\mathrm{F}l$ increases as the resistivity decreases, from 2.4  for the most, to 8.6 for the least disordered film. The elastic scattering length ranges from $2.4\unit{\AA}$ to $7.3\unit{\AA}$, and is of the order of the interatomic distance ($a \approx 4.1\unit{\AA}$ for TiN~\cite{Allmaier:2009fe}). 

Coplanar waveguide quarter-wave resonators were patterned from the films using e-beam lithography and reactive ion etching. The resonators are capacitively coupled to an on-chip feed line, which is wire bonded to coaxial connectors. The sample is mounted onto the cold finger of a He-3 sorption cooler with a base temperature of $310\unit{mK}$. Microwaves from a vector network analyzer are fed to the sample through coaxial cables that are attenuated and filtered at $4\unit{K}$. The amplified forward transmission $S_{21}$ of the feed line is recorded as a function of temperature and microwave frequency. From this transmission spectrum, the resonance frequency of the resonator is determined~\cite{SUPINF}.

\begin{figure}[ht]
\centering
\includegraphics[width=.8\linewidth]{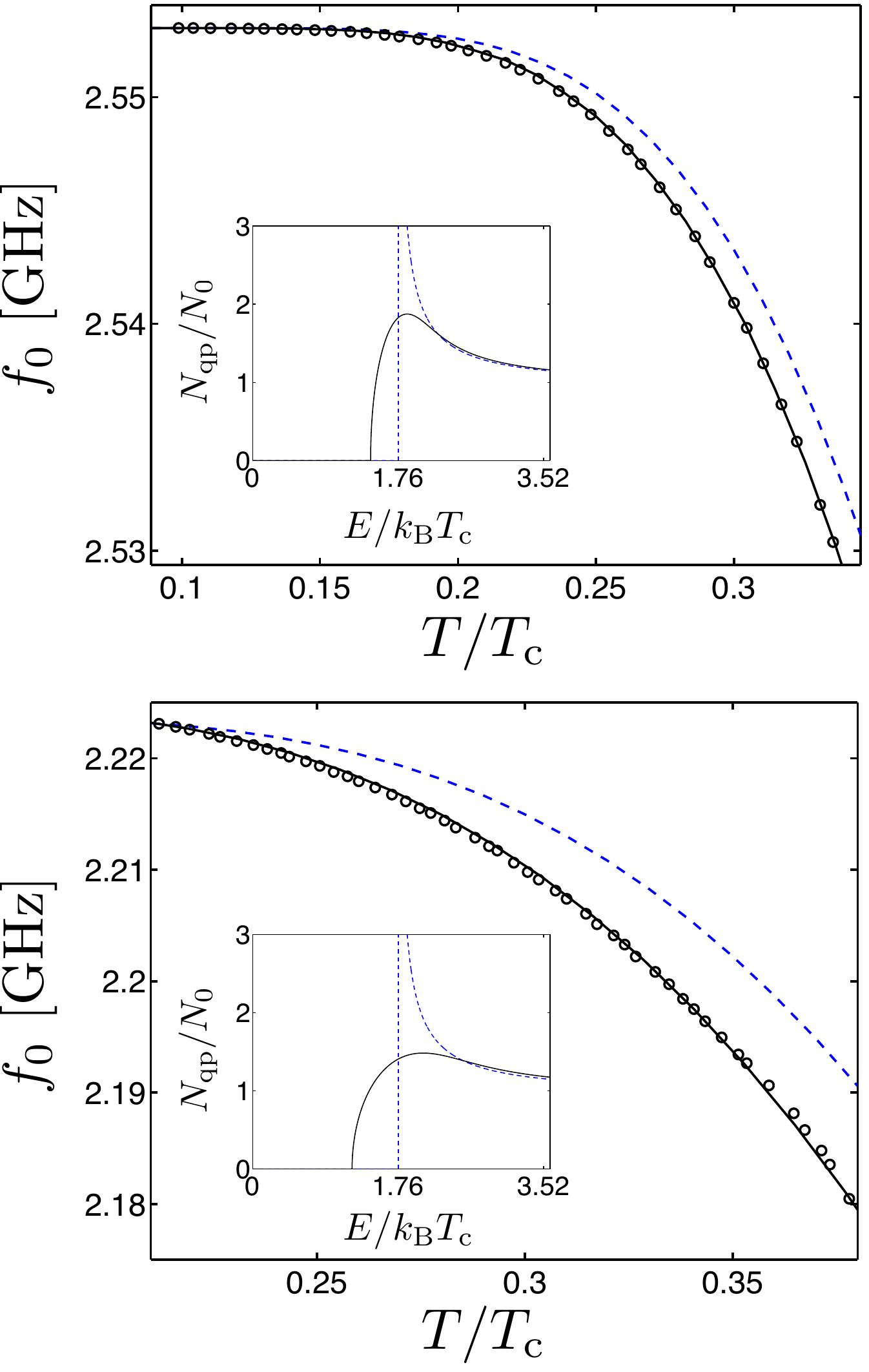}
\caption{(Color online) Measured resonance frequencies as a function of temperature, for films D (top) and A (bottom). The dashed curves are calculated resonance frequencies according to BCS theory. The solid curves are a fit using a broadened density of states (see text for details). The insets show the quasiparticle densities of states used for calculating the resonance frequencies. The curves for all films listed in Table~\ref{tab:parameters} are presented in the supplementary information~\cite{SUPINF}.}\label{fig:resonance}
\end{figure}

%
%

Figure~\ref{fig:resonance} shows two typical traces of the measured resonance frequency $f_0$ (circles) as a function of temperature, for the resonators made from films D and A. The resonance frequency decreases with increasing temperature, which reflects a weakening of the superconducting state. Since our films are in the local limit $l\ll\xi$, with $\xi$ the superconducting coherence length, we assume that the response to microwaves can be described by a complex conductivity $\sigma = \sigma_1 - i\sigma_2$. The resonance frequency of the resonators is a function of $\sigma_2$ and geometric factors only~\cite{SUPINF}. The dashed curves are predictions of this temperature dependence using the complex conductivity as given by Mattis and Bardeen~\cite{Mattis:1958vo}. To obtain these curves, only the resonance frequency at the lowest measured temperature was taken as a scaling parameter~\cite{SUPINF}. It is clear from this comparison, that the measured resonance frequency decays faster with temperature than predicted, as was already noticed in Ref.~\cite{Barends:2008fg} for a single NbTiN film. Moreover, the deviations from theory become stronger for increasing disorder.
We attribute these deviations to an intrinsic, disorder-induced change in the superconducting state. In the remainder of this Letter, we will analyze the observed deviations in more detail, and compare them with the available theory for strongly disordered superconductors.


The effective attraction between electrons in a superconductor is the result of a competition between an attractive electron-phonon interaction and a repulsive Coulomb interaction~\cite{ALLEN:1975ud}.
It is now known that disorder can change the critical temperature of a superconducting film. This change was first attributed to changes in the electron-phonon coupling
 ~\cite{Keck:1976kn}, but later it has become clear that disorder has a strong effect on the effective Coulomb repulsion.
  This insight was used by Finkelstein~\cite{Finkelshtein:1987tl} to explain the decrease in $T_\textrm{c}$ of two-dimensional~\red{\cite{footnote3}} MoGe films with increasing sheet resistance $R_\Box=\rho/d$. In this framework, the critical temperature is given by
\begin{equation}
\Tc = \frac{\hbar}{\kB\tau}\left(\frac{\sqrt{g}-\sqrt{g_\mathrm{c}} \red{+} \sqrt{g_\mathrm{c}/8\pi g}}{\sqrt{g}+\sqrt{g_\mathrm{c}} \red{+} \sqrt{g_\mathrm{c}/8\pi g}}\right)^{\sqrt{\pi g /2},}\label{eq:finkelstein}
\end{equation}
where $g = 2\pi\hbar/e^2R_\Box$ is the dimensionless film conductance, and $g_\mathrm{c} = \ln^2(\hbar/k_\mathrm{B}T_\mathrm{c0}\tau)/2\pi$ is a critical conductance, with $T_\mathrm{c0}$ the critical temperature of the non-disordered material. This mechanism, where superconductivity is destroyed due to the disappearance of attractive interaction, is dubbed the fermionic mechanism, in contrast to the bosonic mechanism where phase fluctuations between weakly connected superconducting islands destroy global coherence. We do not include phase fluctuations in our current analysis, since we expect that these will mostly influence the superconducting state close to $\Tc$, and close to the superconductor-to-insulator transition, i.e. far from our experimental situation~\cite{Mondal:2011da}.

Equation~(\ref{eq:finkelstein}) describes the critical temperature of a uniform superconductor. The short elastic scattering length however will enhance localization, and will lead to mesoscopic fluctuations of the pairing potential $\Delta = \Delta(E, \overrightarrow{r})$~\cite{Skvortsov:2005eb}. The average value of the pairing potential, $\langle\Delta\rangle$, still follows the BCS temperature dependence.  Fluctuations of $\Delta$ impact the quasiparticle density of states however~\cite{Feigelman:2011vj} and can be described on a mean field level by introducing a parameter $\eta$ in the Usadel equation~\cite{Larkin:1972vn}
\begin{equation}
iE\sin\theta + \langle\Delta\rangle\cos\theta - \langle\Delta\rangle\eta\sin\theta\cos\theta = 0,\label{eq:usadel}
\end{equation}
where $E$ is the energy taken from the Fermi energy, $\theta$ the pairing angle, and $\sin\theta$ and $\cos\theta$ are the quasi-classical, disorder-averaged Green's functions. The parameter $\eta$ is a measure for the spatial correlations in $\Delta(r)$. Assuming a 3-dimensional superconductor and long-range spatial correlations, its temperature-dependent value is given by~\cite{Feigelman:2011vj}
\begin{equation}
\eta(T/T_\mathrm{c}) \approx \frac{K(T/T_\mathrm{c})}{4g(g-g_\mathrm{c})},\label{eq:eta}
\end{equation}
where $K(T/\Tc)$ is a universal function of temperature, with $K(0) = 1$.

After solving Eq.~(\ref{eq:usadel}), the quasiparticle density of states is given by $N_\mathrm{qp} = N_0~\mathrm{Re}(\cos\theta)$, with $N_0$ the density of states in the normal metal. For increasing values of $\eta$, the coherence peak at $E=\Delta$ is lowered and broadened, as qualitatively ilustrated in the insets of Fig.~\ref{fig:resonance}. A finite density of states appears for energies $E<\langle\Delta\rangle$. In the mean-field description used here, for low enough values of $\eta$, the density of states stays gapped. Localized states will however create a finite but small density of sub-gap states~\cite{Balatsky:2006ce,Feigelman:2011vj}. This so-called Lifshitz tail will not be taken into account in the rest of this Letter.


The imaginary part of the conductivity can be calculated using a generalized Mattis-Bardeen equation~\cite{Nam:1967ue}, that is valid for a superconductor with arbitrary pairing angle $\theta$ describing the extended electronic states:
\begin{eqnarray}
\hbar\omega\frac{\sigma_2}{\sigma_\mathrm{n}} = & \int_{-\hbar\omega}^{\infty}\mathrm{d}E~g_2(E,E')\left[1-2f(E')\right]\nonumber\\
&+\int_0^\infty\mathrm{d}E~g_2(E',E)\left[1-2f(E)\right]\label{eq:sigma2},
\end{eqnarray}
with $\omega$ the microwave frequency, $\sigma_\mathrm{n}$ the normal state conductivity, $E' = E + \hbar\omega$, $f(E)$  the quasiparticle distribution function, and the generalized coherence factor $g_2(E,E')  = \mathrm{Re}[\cos\theta(E)]\mathrm{Im}[\cos\theta(E')]+\mathrm{Im}[i\sin\theta(E)]\mathrm{Re}[i\sin\theta(E')]$. For $\eta=0$, Eq.~(\ref{eq:sigma2}) reduces to the standard Mattis-Bardeen expression.
Qualitatively, a broadened density of states as derived from this theory, can be used to describe the resonator measurements presented here. Quantitatively, the comparison fails however. Even for film A, where the broadening effects are expected to be largest, only a value of $\eta\approx 10^{-4}$ is predicted from Eq.~(\ref{eq:eta}), whereas a value of $\eta \approx 0.1$ would be needed to describe our measurements.

To make further progress in analyzing the data, we modify the approach taken by Feigel'man and Skvortsov, by assuming an arbitrary pair breaking parameter $\alpha$ in the Usadel equation (\ref{eq:usadel}), which then reads
\begin{equation}
iE\sin\theta + \Delta\cos\theta - \alpha\sin\theta\cos\theta = 0.
\end{equation}
This parameter describes a spatially uniform breaking of time-reversal symmetry, the origin of which needs to be determined. 
 We also assume that, in contrast to the model discussed above, this pair breaking mechanism influences the magnitude of the pairing amplitude $\Delta$ via the self-consistency equation
\begin{equation}
\Delta = N_0V \int_0^{k_\textrm{B}\Theta_\textrm{D}}\textrm{d}E~\textrm{Im}(\sin\theta)[1-2f(E)],\label{eq:delta}
\end{equation}
where $V$ is a measure for the effective interaction of the undisturbed system, and $\Theta_\mathrm{D}$ is the Debye temperature. Both $V$ and $\Theta_\mathrm{D}$ determine the critical temperature $\Tc'$ for $\alpha\rightarrow0$. For each value of $\alpha$ we \red{\sout{identify }determine} the temperature where $\Delta\rightarrow0$\red{, and we identify this with\sout{ as}} the \red{measured} critical temperature $\Tc$.
This description is analogous to the description of Abrikosov and Gor'kov for the effect of magnetic impurities~\cite{ABRIKOSOV:1961uo}, or to the description of a superconducting wire carrying a current~\cite{Anthore:2003fm}.

\begin{figure}[ht]
\centering
\includegraphics[width=.8\linewidth]{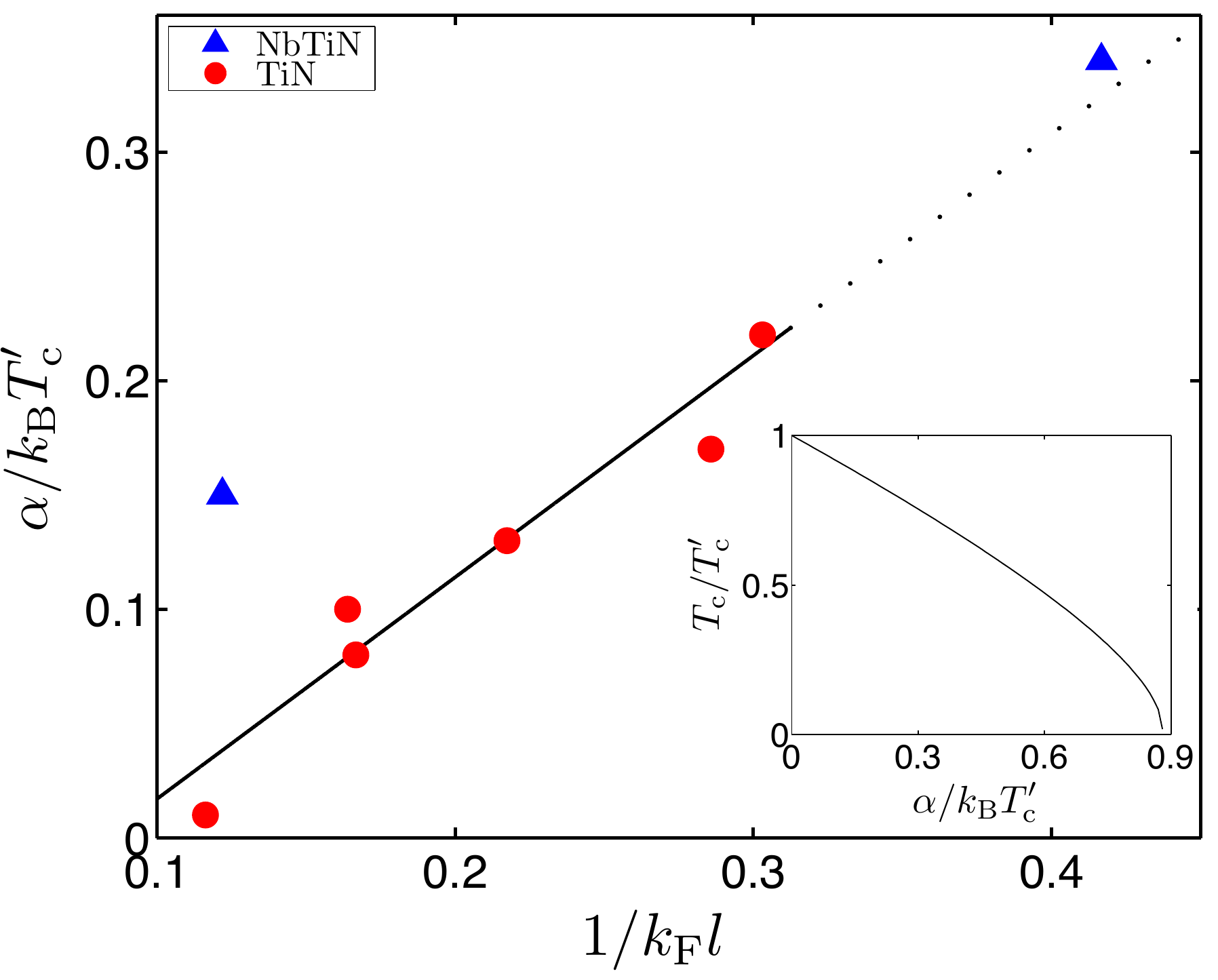}
\caption{(Color online) Value of the pair breaking parameter $\alpha$, extracted from fits of the resonance frequency, versus disorder parameter $1/k_\mathrm{F}l$, for the series of TiN films (circles) and the two NbTiN films (triangles). The line is a linear fit to the TiN data. The inset shows the calculated reduction of $\Tc$ as a function of $\alpha$.}\label{fig:alpha}
\end{figure}

We use this analysis for the measured resonance frequencies, as shown in Fig.~\ref{fig:resonance} (solid curves). We take again the measured resonance frequency at the lowest temperature as a scaling parameter, and then we adjust $\alpha$ to find the best fit. This procedure implicitly varies the value for $\Tco$ for each value of $\alpha$. The density of states (at $T=0$) that corresponds to the fitted value of $\alpha$ is shown in the inset. We find that the value of $\alpha$ increases with the level of disorder $1/k_\mathrm{F}l$, as is shown in Fig.~\ref{fig:alpha}. For the series of TiN films (circles), $\alpha$ has a linear relation to $1/k_\mathrm{F}l$. For the two NbTiN films (triangles), the trend is similar: increasing disorder leads to a higher value of $\alpha$. 

\begin{figure}[ht]
\centering
\includegraphics[width=.8\linewidth]{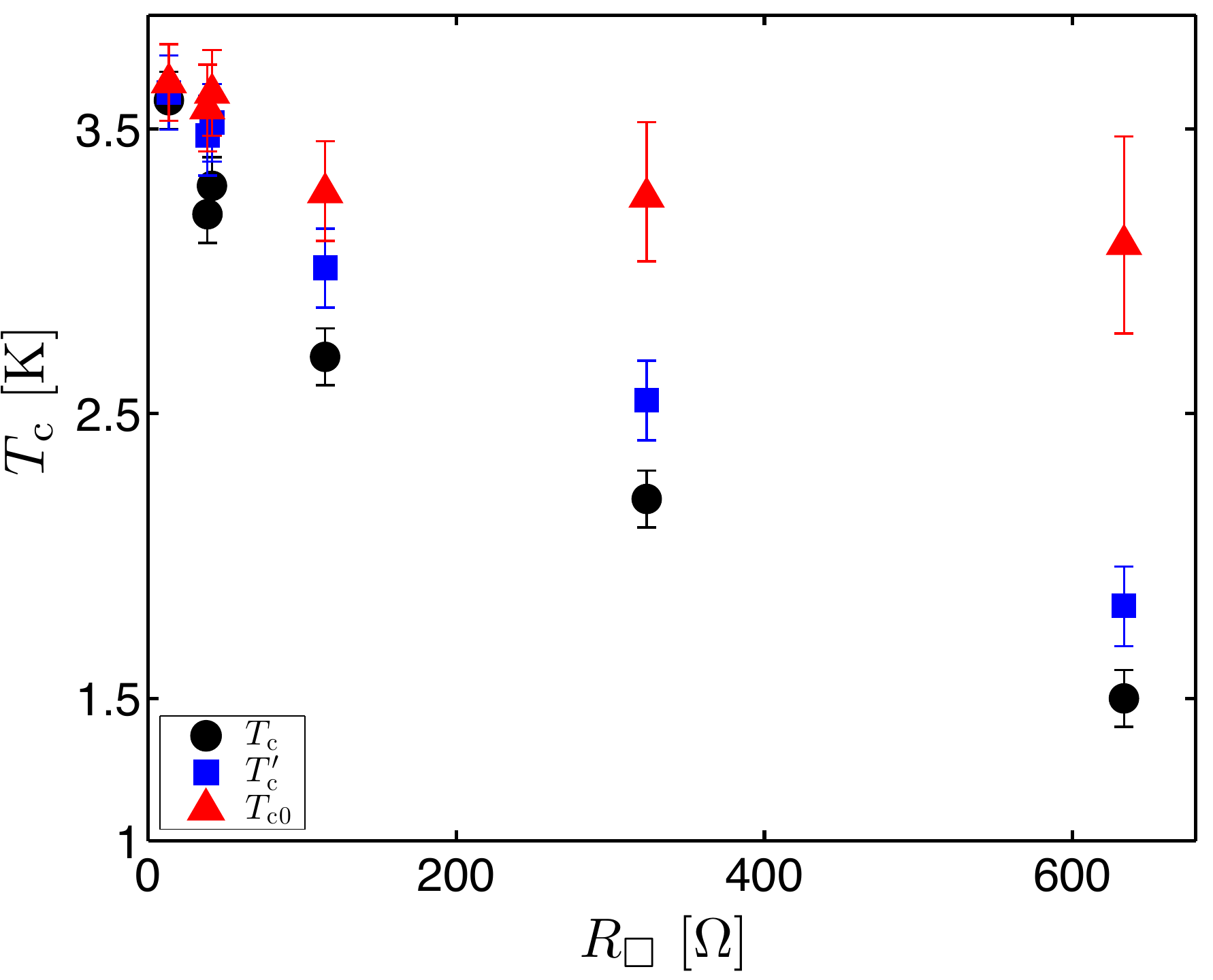}
\caption{(Color online) Measured critical temperature $\Tc$ as a function of sheet resistance $\Rsq$ (circles), for the series of TiN films A-F. Squares show the value of $\Tc'$, when the reduction due to $\alpha$ is accounted for. Triangles give the value for $\Tco$ when  reduction due to Coulomb interaction [see Eq.~(\ref{eq:finkelstein})] is also taken into account.}\label{fig:tc}
\end{figure}

As a consistency check, we analyze the critical temperatures of the TiN films. The pair breaking parameter $\alpha$ will reduce the critical temperature, as shown in the inset of Fig.~\ref{fig:alpha}. The measured critical temperature of films A-F decreases with sheet resistance, as shown in Fig.~\ref{fig:tc} (circles). The squares in Fig.~\ref{fig:tc} show the critical temperature when corrected for the effects of $\alpha$. Increased Coulomb interaction also reduces the critical temperature, as described by Eq.~(\ref{eq:finkelstein}). Taking both reductions into account, using a constant value of $\tau=1.2\unit{fs}$ (see Table~\ref{tab:parameters}), we find a value of $\Tco$ that is constant within 15\% (triangles in Fig.~\ref{fig:tc}). This is consistent with the self-limiting character of the ALD deposition process, from which we can expect that all films are metallurgically identical.

In our analysis, we have only taken  into account modifications of extended electronic states. Localized states will however be present, and will generate a finite density of states in the quasiparticle gap~\cite{Balatsky:2006ce,Feigelman:2011vj}. The effect of these localized states will be most pronounced at low temperatures, where mostly sub-gap states will be thermally occupied.  \red{At temperatures $T<0.15\ \Tc$ we observe a slower saturation of the resonance frequency than expected from our analysis. We tentatively attribute this to the presence of localized states.\sout{stronger temperature dependence than predicted by our analysis}. A description including these states will necessarily go beyond the description of the electrodynamics by Nam~\cite{Nam:1967ue}. At these temperatures, the relative change in resonance frequency is of the order $<10^{-4}$. The small deviations therefore do not affect our analysis at higher temperatures.}


Let us now consider the possible origins of the pair breaking parameter $\alpha$. From the fact that the resonance frequency decreases monotonically with temperature, and the fact that the measured deviations are comparable for different substrates and different films, we do believe that the role of magnetic moments in the dielectric surroundings of our resonator is negligible~\cite{Barends:2008fg}. We can also exclude the effect of an uncontrolled level of magnetic impurities. There is a very small concentration of iron expected in our films (the TiCl$_4$ precursor gas contains 0.12 mass-ppm of Fe), but since we expect all TiN films to be metallurgically identical, this small fraction of magnetic impurity would yield a constant value for $\alpha$, contrary to the trend that is observed. A quantitative explanation of our results might be sought along the lines of disorder-induced spin fluctuations~\cite{Finkelstein:1984up,Ebisawa:1985cl}. Alternatively, the small elastic scattering length $l$ might enhance the effect of a minute fraction of magnetic impurities~\cite{BHATT:1992br}. In both cases, the effects on the superconducting state are primarily caused by the level of disorder, and are intrinsically present in the materials.

In conclusion, we have shown that the superfluid part of the complex conductivity, $\sigma_2$, as probed with microwave resonators, increasingly deviates from conventional Mattis-Bardeen theory for thin films with normal state resistivities exceeding $100\unit{\mu\Omega cm}$.  We have analyzed these results with the conjecture of a disorder-dependent change of the density of states, including a lowering of the coherence peak and the occurrence of states for energies below the pairing potential $\Delta$, as suggested by  theoretical predictions~\cite{Ghosal:1998up} and previous experimental results~\cite{Sacepe:2008jx}. Using a heuristic approach, inspired by Feigel'man and Skvortsov~\cite{Feigelman:2011vj}, we provide a consistent interpretation of the observed electrodynamics of films with these high resistivities.  Although further details need to be clarified, these observations underline that disorder has a much more pronounced effect on the microscopic electronic properties than is often assumed for s-wave superconductors.

\begin{acknowledgments}The authors thank Nathan Vercruyssen, Misha Feigel'man,  Misha Skvortsov, and Jochem Baselmans for helpful discussions, and Marc Zuiddam and Tony Zijlstra for help in depositing the various thin films.  TMK thanks the Keck Institute for Space Studies for supporting his stay at California Institute of Technology, which helped to initiate this work. The research was financially supported by MicroKelvin (No. 228464, Capacities Specific Programme) and the Dutch Foundation for Research of Matter (FOM).\end{acknowledgments}

%
\newpage
\onecolumngrid
\section{Supplementary material for ``Strongly disordered s-wave superconductors probed by microwave electrodynamics''}

\subsection{E. F. C. Driessen, P. C. J. J. Coumou, R. R. Tromp, P. J. de Visser, and T. M. Klapwijk}

\author{E. F. C. Driessen}
\email{e.f.c.driessen@tudelft.nl}
\affiliation{Kavli Institute of Nanoscience, Delft University of Technology, Lorentzweg 1, 2628 CJ Delft, The Netherlands}

\author{P. C. J. J. Coumou}
\affiliation{Kavli Institute of Nanoscience, Delft University of Technology, Lorentzweg 1, 2628 CJ Delft, The Netherlands}

\author{R. R. Tromp}
\affiliation{Kavli Institute of Nanoscience, Delft University of Technology, Lorentzweg 1, 2628 CJ Delft, The Netherlands}

\author{P. J. de Visser}
\affiliation{Kavli Institute of Nanoscience, Delft University of Technology, Lorentzweg 1, 2628 CJ Delft, The Netherlands}
\affiliation{SRON National Institute for Space Research, Sorbonnelaan 2, 3584 CA Utrecht, The Netherlands}

\author{T. M. Klapwijk}
\affiliation{Kavli Institute of Nanoscience, Delft University of Technology, Lorentzweg 1, 2628 CJ Delft, The Netherlands}

\maketitle

\section{Film deposition details}

Films A-F and H were grown on high-resistivity ($\rho > 1\unit{k\Omega cm}$) Si(100) substrates. Substrates F and H were hydrogen passivated by dipping  in hydrofluoric acid prior to the deposition. In the case of films A-E, this treatment was not performed and there was a thin native layer of silicon oxide present underneath the film. Film G was grown on A-plane sapphire. 

The titanium nitride films were grown using plasma-assisted atomic layer deposition (ALD)~\cite{Heil:2007hy}. In this process, TiCl$_4$ and a plasma of H$_2$ and N$_2$ are stepwise repetitively introduced into an Oxford Instruments reaction chamber and react into TiN and gaseous HCl. After each step, the residual gas is removed from the reaction chamber by an argon gas purge. During the deposition, the substrate is heated to 400~$^\circ$C, while the walls of the reaction chamber are kept at 80~$^\circ$C. The average growth rate of $0.45\unit{\AA}$ was determined with a step profiler on thick films.

The NbTiN films F and G were deposited using DC magnetron sputtering, at a DC power of $300\unit{W}$, using an argon flow of $100\unit{sccm}$ and a nitrogen flow of $4\unit{sccm}$. The pressure in the chamber was varied to change the resistivity of the resulting film ($4\unit{mTorr}$ for film G, $12\unit{mTorr}$ for film H)~\cite{Iosad:2000gz}.

\section{Determination of material parameters}
The thickness $d$ of films B-E is determined with a step profiler on a co-deposited lift-off structure. The thicknesses of the other films was estimated from known deposition rates.

The grain size $b$ was determined from an atomic force micrograph of the surface of the films. To estimate the grain size, the diameter of a few representative grains was determined and averaged.

To determine the electric parameters of the films, we measured sheet resistance $\Rsq$, Hall voltage $V_\mathrm{H}$, and (for selected films) upper critical field $H_\mathrm{c2,\perp}$  for the films in Table I of the main article. From these measurements, resistivity $\rho$, carrier density $n$, and diffusion constant $D$ were directly determined:
\begin{eqnarray}
\rho &= &\Rsq d,\\
n & = &\frac{IB}{V_\mathrm{H}ed},\\
D & = &\frac{4\kB}{\pi e}\left(\frac{dH_\mathrm{c2,\perp}}{dT}\right)^{-1}_{T=\Tc},
\end{eqnarray}
with $d$ the film thickness, $I$ and $B$ applied DC current and magnetic field, respectively, and $T$ the temperature.

To estimate values for the different listed parameters, we use the Drude-Sommerfeld model for the resistivity~\cite{Ashcroft:1976wl}
\begin{equation}
\rho = \frac{m^\ast}{n e^2 \tau},\label{eq:drude}
\end{equation}
with $m^\ast$ the effective mass of the charge carriers, and $\tau$ their relaxation time. Furthermore, we assume relaxation by elastic scattering such that $\tau = l / v_\mathrm{F}$, with $l$ the elastic scattering length, and $v_\mathrm{F}$ the Fermi velocity.

The Fermi wave vector is estimated from the carrier density $n$, assuming again free electrons,
\begin{equation}
k_\mathrm{F} = \sqrt[3]{3\pi^2 n},
\end{equation}
and the Fermi velocity and wave vector are related through
\begin{equation}
v_\mathrm{F} = \frac{\hbar k_\mathrm{F}}{m^\ast}.
\end{equation}
Combining these equations, a value for $l$ and $k_\mathrm{F}l$ can be estimated as listed in Table~I, from resistivity and carrier density only.

To estimate the scattering time, the knowledge of $l$ is combined with the diffusion constant that is measured from the upper critical field. In a 3-D geometry, it is given by $D = v_\mathrm{F}l/3$, yielding
\begin{equation}
\tau = \frac{l^2}{3D}.
\end{equation}

The critical temperature was determined from a measurement of the resistive transition. The resistive transition is smooth for all films. Therefore, the temperature at which the resistance was reduced by 10\% was taken as an upper bound of the critical temperature. For most films, the measurement accuracy of $\Tc$ was 0.1~K due to a resolution limit in the thermometer used.

\section{Resonance frequency of a coplanar waveguide superconducting resonator}
For a thin superconducting film in the dirty limit, with $\sigma = \sigma_1 - i\sigma_2$ and $\sigma_2 \gg \sigma_1$, the sheet inductance at angular frequency $\omega$ is given by~\cite{HENKELS:1977ul}
\begin{equation}
L_\mathrm{S} = \mu_0\lambda\coth\frac{d}{\lambda},
\end{equation}
with $\lambda = 1/\sqrt{\mu_0\omega\sigma_2}$ the magnetic penetration depth, and $d$ the film thickness.

The resonance frequency $f_0$ of an uncoupled quarter-wave coplanar waveguide resonator is determined by the kinetic inductance of the superconducting film, and geometric parameters
\begin{equation}
f_0 = \frac{1}{4 l \sqrt{(L_\mathrm{g}+\gamma L_\mathrm{S})C}},
\end{equation}
with $l$ the resonator length, $L_\mathrm{g}$ the geometric inductance, $\gamma$ a geometric parameter converting sheet inductance to kinetic inductance, and $C$ the geometric capacitance per unit length. The resonator length and geometric capacitance are mere prefactors in this equation, and in our comparison, they are incorporated in a scaling factor at the lowest temperature.

\begin{table}[ht]
\centering
\caption{Geometric parameters of the films used in the analysis of our experiments.}\label{tab:geometric}
\begin{tabular}{cccccc}
\hline\hline
Film	& $d$	& $S$		& $W$		& $L_\mathrm{g}$	& $\gamma$\\
	& [nm]	& [$\unita{\mu m}$]	& [$\unita{\mu m}$]	& [$\unita{nH / m}$]		& [$\unita{mm^{-1}}$]\\
\hline
A	& 6		& 10			& 2			& 311			& 288\\
B	& 11		& 10			& 2			& 311			& 273\\
C	& 22		& 10			& 2			& 311			& 256\\
D	& 45		& 10			& 2			& 311			& 238\\
E	& 89		& 10			& 2			& 311			& 220\\
F	& 55		& 3			& 2			& 437			& 482\\
\hline
G	& 300	& 3			& 2			& 437			& 391\\
H	& 50		& 3			& 2			& 437			& 486\\
\hline\hline
\end{tabular}
\end{table}

The geometric inductance for a coplanar waveguide geometry with central line width $S$ and gap width $W$, is given by
\begin{equation}
L_\mathrm{g} = \frac{\mu_0}{4}\frac{K(k')}{K(k)},
\end{equation}
where $K$ is the complete elliptic integral of the first kind, $k = S / (S+W)$, $k^2 + k'^2 = 1$. The geometric parameter $\gamma$ is calculated with~\cite{Holloway:1995ww}
\begin{equation}
\gamma = \frac{1}{32K^2(k)}\frac{(S+2W)^2}{W(S+W)}\left[\frac{2}{S}\ln\left(\frac{S}{\delta}\frac{W}{S+W}\right)+\frac{2}{S+2W}\ln\left(\frac{S+2W}{\delta}\frac{W}{S+W}\right)\right],
\end{equation}
with $\delta = d / [4\pi\exp(\pi)]$. Table~\ref{tab:geometric} lists the values for the different geometric parameters that were used in our analysis.

\begin{figure}[ht]
\centering
\includegraphics[width=.45\linewidth]{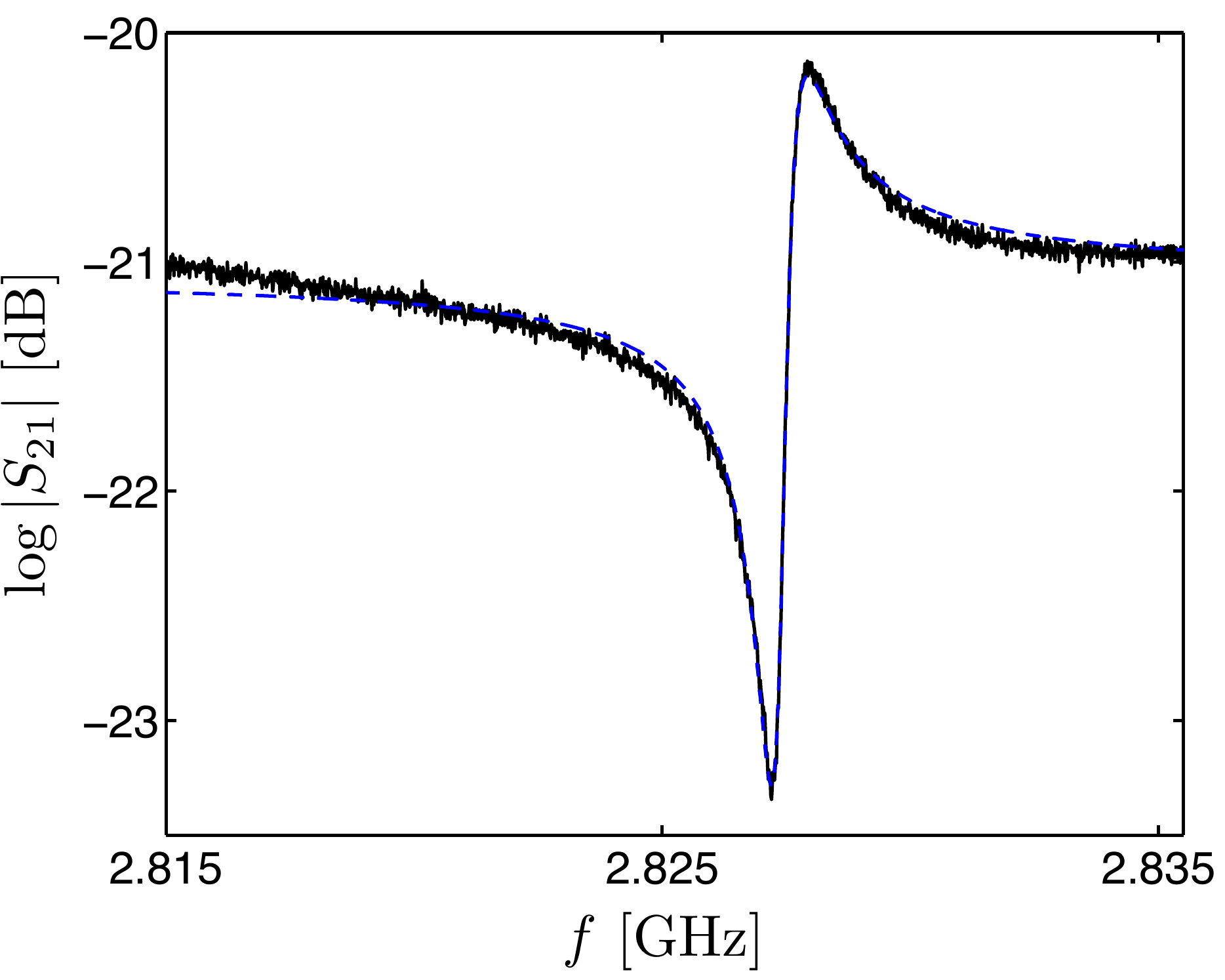}
\caption{Resonance curve of a typical resonator. The dashed curve is a fit to the line shape [Eq.~(\ref{eq:resonance})] used to extract the resonance frequency $f_0$.}\label{fig:rescurve}
\end{figure}

The quarterwave resonator is capacitively coupled to a coplanar waveguide transmission line, of which the power transmission $S_{21}$ is measured as a function of frequency. Figure~\ref{fig:rescurve} gives a typical resonance curve. The resonant line shape is asymmetric due to an impedance mismatch of the transmission line to the $50\unit{\Omega}$ coaxial cables. The resonance frequency $f_0$ is extracted from this line shape by fitting the relation~\cite{Khalil:2012jr}
\begin{equation}
S_{21} = 1-\frac{Q/Q_\mathrm{e}}{1+2iQ\frac{f-f_0}{f_0}},\label{eq:resonance}
\end{equation}
where $Q$ is the loaded quality factor of the resonance, $Q_\mathrm{e}$ is the complex coupling quality factor, and $f$ is the microwave frequency. A small linear background transmission was accommodated in the fit to account for less-than-perfect calibration of the cables. The dashed curve in Fig.~\ref{fig:rescurve} is a fit of this line shape to the shown resonance.

\section{Data from other films}
\begin{figure}[ht]
\centering
\includegraphics[width=.8\linewidth]{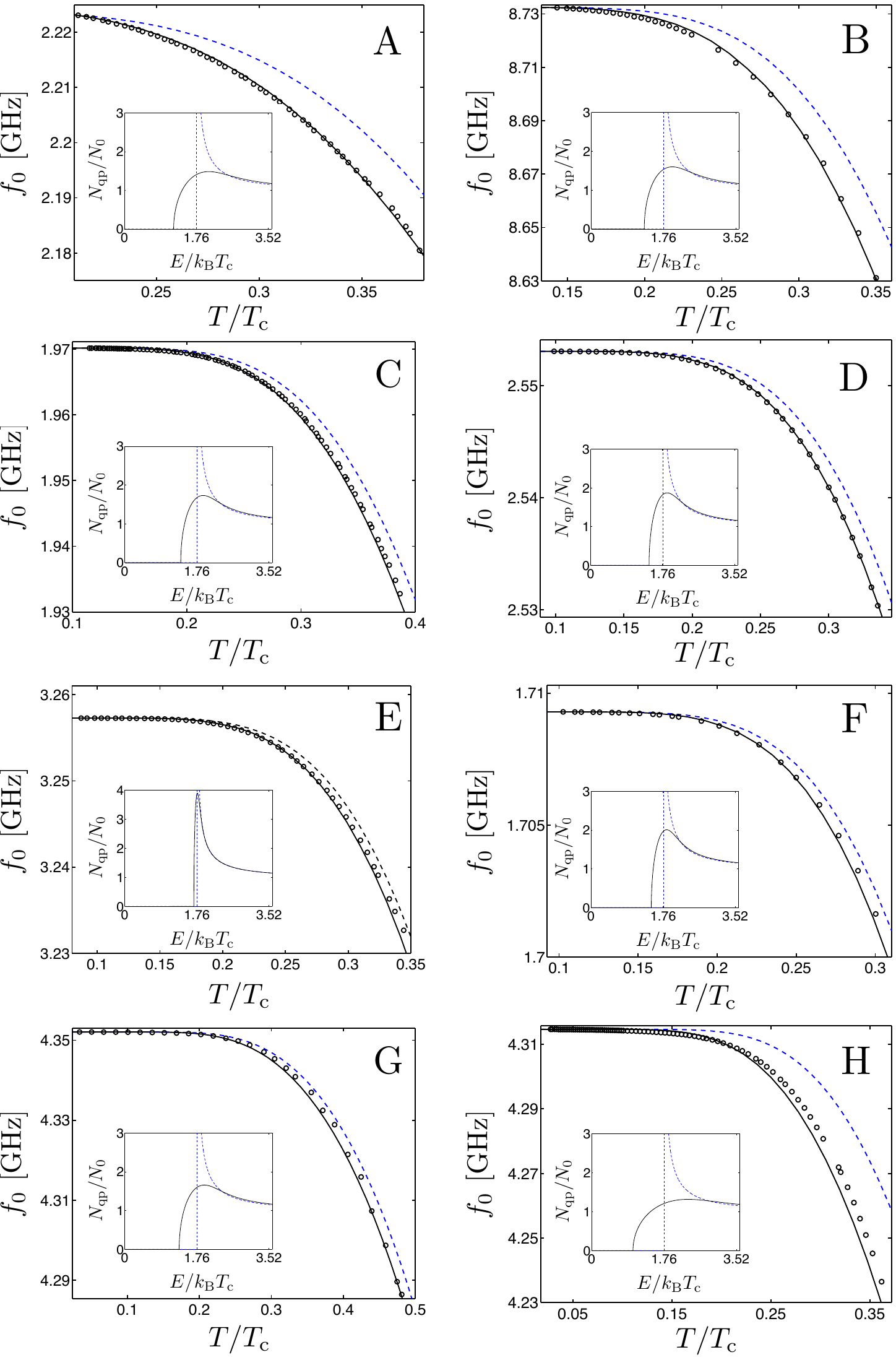}
\caption{Measured resonance frequencies as a function of temperature, for all films. The dashed curves are calculated resonance frequencies according to Mattis-Bardeen theory. The solid curves are a fit using a broadened density of states (see main text for details). The insets show the densities of states used for calculating both curves.}\label{fig:curvesall}
\end{figure}

Figure~\ref{fig:curvesall} shows the measured resonance frequencies for all films presented in the paper. The dashed curves are BCS predictions of the temperature dependence, the solid curves are fits using the model described in the paper. We note here, that decreasing the geometric inductance $L\mathrm{g}$ would shift the calculated Mattis-Bardeen curves towards the measurements. This shift is however much smaller than the observed deviations, even when the geometric inductance is ignored. Moreover, this would not account for the observed trend with disorder.
%

\end{document}